\documentclass[prd,aps,showpacs,tightline,twocolumn,nofootinbib]{revtex4}
\usepackage{graphicx}
\usepackage{amssymb}

\newcommand{\bear}{\begin{array}}  \newcommand{\eear}{\end{array}}
\newcommand{\bea}{\begin{eqnarray}}  \newcommand{\eea}{\end{eqnarray}}
\newcommand{\beq}{\begin{equation}}  \newcommand{\eeq}{\end{equation}}
\newcommand{\bef}{\begin{figure}}  \newcommand{\eef}{\end{figure}}
\newcommand{\bec}{\begin{center}}  \newcommand{\eec}{\end{center}}
\newcommand{\non}{\nonumber}

\newcommand{\del}{\partial}

\newcommand{\la}{\left\langle} \newcommand{\ra}{\right\rangle}

\begin{document}

\title{$Q$-ball Metamorphosis}
\author{Masahiro Kawasaki}
\author{Fuminobu Takahashi}
\affiliation{Research Center for the Early Universe, University 
  of Tokyo, Bunkyo-ku, Tokyo 113-0033, Japan}
\date{\today}

\begin{abstract}
Flat directions in the minimal supersymmetric standard model are known
to deform into non-topological solitons, $Q$-balls, which generally
possess both baryon and lepton asymmetries. We investigate how
$Q$-balls evolve if some of the constituent fields of the flat
direction decay into light species. It is found that the $Q$-balls
takes a new configuration whose energy per charge 
slightly increases due to the decay.  
Specifically, we show that all the stable $Q$-balls eventually transform into pure $B$-balls via the decay into neutrinos.
\end{abstract}
\pacs{98.80.Cq \hspace{8.0cm} RESCEU-7/04}
\maketitle

\section{introduction} 
\label{sec:intro}

Non-topological solitons, such as $Q$-balls~\cite{Coleman:1985ki} and
$I$-balls~\cite{Kasuya:2002zs}, play important roles in the particle
cosmology, partly because they are (quasi-)stable objects. Their
stability comes from the conservation law. The $Q$-ball is composed of
a complex scalar field with an $U(1)$ charge, the conservation of
which enables the $Q$-ball to be stable or long-lived. Similarly,
$I$-balls are composed of a real scalar field (or a complex scalar
field in a nearly straight-line motion), whose dynamics conserve the
adiabatic charge.  Since we can only see either these objects in their
final state or their decay products in the present universe, it is 
important to study their evolution during the course of the expanding
universe.

The baryon asymmetry is one of the key observational results to
uncover the history of the universe. Among many baryogenesis scenarios
proposed so far, the mechanism proposed by Affleck and
Dine~\cite{Affleck:1984fy}  can be realized in the minimal
extension of the standard model, that is, the minimal supersymmetric
standard model (MSSM).  The mechanism makes use of one of flat
directions, along which there is no classical potential
in the supersymmetric limit.  Although the
flat direction is parametrized by a set of chiral superfields, it is
convenient and even sufficient in the usual situation to express it in
terms of a single complex scalar field dubbed `Affleck-Dine (AD)
field,' $\Phi$, which generally has nonzero baryon and lepton
charges. The dynamics of the AD field not only generates the desired
baryon asymmetry, but allows non-topological solitons to be formed.
The properties of $Q$-balls in the context of the AD mechanism have
been extensively
studied~\cite{Kusenko:1997si,Enqvist:1998en,Kasuya:1999wu,
Kasuya:2000wx}.  $Q$-balls
can be stable or unstable depending on the situation. Stable $Q$-balls
can be dark matter~\cite{Kusenko:1997si,Kasuya:2000sc,
Kasuya:2001hg}, which might
solve the coincidence problem between the abundance of baryons and
that of dark matter.  Unstable but long-lived $Q$-balls can keep the
baryon and lepton asymmetries stored inside them from being subject to
the sphaleron effects.  For instance, they can successfully generate
both large lepton asymmetry and small baryon
asymmetry~\cite{Kawasaki:2002hq}. It is even possible to realize the
late-time baryon appearance after the relevant BBN epoch with use of
$Q$-balls~\cite{Ichikawa:2004pb}. $I$-balls are also formed in the AD
mechanism as an excited state of $Q$-balls.

In the examples mentioned above, the single-field parametrization is
simple and valid. However, if we examine the evolution of $Q$-balls
closely, we have to pay much more attention to the fact that the flat
direction is actually composed of several scalar fields.  Indeed, the
slepton condensate always decays into a pair of neutrinos, while the
squark condensate can decay into hadrons only if its energy per one
baryon charge exceeds the nucleon or meson masses. 
Therefore, if a $Q$-ball
has nonzero baryon and lepton asymmetries and its energy per unit
charge is small enough, the leptonic component decays, leaving only
the baryon asymmetry inside $Q$-balls ({\it i.e.}, pure $B$-balls).
However, since the $Q$-ball solution is obtained under the
assumption that the $U(1)$ charge, $Q$, is conserved, it is not trivial
how the $Q$-ball configuration changes during the course of the
partial decay.  Do $Q$-balls come apart? Or do they change their shape
to satisfy the modified $Q$-ball solution?  It is the purpose of this
study to answer these questions.

In this paper, we investigate how $Q$-balls evolve if some of the
constituent fields decay into lighter species. In the next section, we
review the flat direction of the MSSM and the $Q$-ball solution. In
section~\ref{sec:model} we study the partial decay of $Q$-balls with a
toy model of the flat direction.  The results obtained there are
confirmed with use of numerical calculation presented in
section~\ref{sec:nume}.  Finally we give discussions and conclusions
in section~\ref{sec:discuss}.

\section{Preliminaries} 
\label{sec:preli}
In this section, first we explain how flat directions are expressed,
paying particular attention to the single-field parametrization.  Then
we review the $Q$-ball solution relevant for the next section.

\subsection{Flat directions}
In the MSSM, there are many flat directions along which both the
$D$-term and $F$-term potentials vanish at the classical level.  The
$D$-flat direction is labeled by a holomorphic gauge-invariant
monomial, $X$. Most of $D$-flat directions can be also $F$-flat,
simply due to a generation structure of the quark and lepton sector.
Since it is easy to satisfy the $F$-flat condition, let us concentrate
on the $D$-flat condition in the following. The $D$-flat direction,
$X$, can be expressed as
\beq
\label{eq:monomial}
X \equiv \prod_{i}^{N} \Phi_i,
\eeq
where $N$ superfields $\left\{ \Phi_i \right\}$ constitute the flat
direction $X$, and we have suppressed the gauge and family indices
with the understanding that Latin letter $i$ contains all the
information to label those constituents.  When $X$ has a non-zero
expectation value, each constituent field also takes a non-zero
expectation value
\begin{equation}
\label{eq:chivev}
\la \Phi_i \ra = \frac{\phi_i}{\sqrt{2}} e^{i \theta_i}.
\end{equation}
Here each $\phi_i/\sqrt{2}$ is the absolute value of the expectation
value $\la \Phi_i \ra$, and is related to every other due to the
$D$-flat conditions:
\beq
\label{eq:D-flat}
\frac{1}{2} \sum_A \left(
\sum_{ij} \Phi_i^* (T_A)_{ij} \Phi_j\right)^2 = 0,
\eeq
where $T_A$ are hermitian matrices representing the generators of the
gauge algebra, and they are labeled with $A$. This condition can be
usually satisfied if we take all $\phi_i$ equal: $\phi_i \equiv
\phi/\sqrt{N}$. Note that $D$-flat condition dictates that all the
amplitudes be equal, while the phases, $\theta_i$, remain to be
arbitrary. In fact, it is necessary to know the pattern of baryon and
lepton symmetry breaking, in order to identify the Nambu-Goldstone
(NG) boson relevant for the AD mechanism. The situation becomes simple
if the spontaneously broken symmetry coincides with the explicitly
violated one. This is the case if the A-term is given by some powers
of $X$. Then all $\theta_i$ become equivalent, and the NG boson can be
identified with the average of $\theta_i$~\cite{Takahashi:2003db}:
\beq
\theta \equiv \frac{1}{N} \sum_i \theta_i,
\eeq
where $\theta$ is orgthogonal to the NG mode corresponding to the
spontaneously broken $U(1)$ gauge symmetry, that is,  the 
linear combination of left isospin $T_{L3}$ and weak hypercharge $Y$.
This is because 
$\sum_i  T_{L3, i}=\sum_i  Y_i=0$
by definition of the $D$-flat direction.
Thus, the dynamics of the flat direction, $X$, can be well described by
one complex scalar field, $\Phi$, which is defined as
\beq
\Phi \equiv \frac{\phi}{\sqrt{2}} e^{i \theta}.
\eeq
Note that $\Phi$ is canonically normalized.

Flat directions are lifted by the supersymmetry (SUSY) breaking effect and
non-renormalizable terms. Since we are interested in the evolution of
$Q$-balls which are formed after the flat direction starts to
oscillate, we neglect the non-renormalizable terms in the following
discussion.  To be concrete, let us adopt the gravity-mediated SUSY
breaking model. The potential of the flat direction is now given by
\bea
\label{eq:potential}
V(\left\{\Phi_i\right\}) &=& \sum_i m_{i}^2 |\Phi_i|^2 \left(
1+K_i \log\left(\frac{|\Phi_i|^2}{M_*^2}\right)\right)\non\\
&&+
\frac{1}{2} \sum_A \left(
\sum_{ij} \Phi_i^* (T_A)_{ij} \Phi_j\right)^2,
\eea
where $m_i$ is a soft mass, $K_i$ a coefficient of the one-loop
correction, $M_*$ the renormalization scale to define the mass.  This
potential reduces to the following in the single-field representation,
\bea
\label{eq:phi_pote}
V(\Phi) &=&  m_{\Phi}^2 |\Phi|^2 \left(
1+K \log\left(\frac{|\Phi|^2}{N M_*^2}\right)\right),
\eea
where $m_\Phi$ and $K$ are defined as
\beq
m_\Phi^2 =\frac{1}{N} \sum_i m_i^2,~~~K= \frac{\sum_i m_i^2 K_i}{\sum_i m_i^2}.
\eeq
As long as the $D$-flat condition is satisfied and all $\Phi_i$ can be
treated equally, the single-field parametrization is useful, and we
only have to deal with the simple potential Eq.~(\ref{eq:phi_pote}).
However, once this assumption breaks down (say, some of the
constituent fields decay), it is necessary to adopt the full potential
Eq.~(\ref{eq:potential}) in order to follow the evolution of each
$\Phi_i$.

\subsection{Q-ball solution}
During inflation, the flat direction, $X$, is assumed to take a large
expectation value $\la X \ra \ne 0$. After inflation ends, the Hubble
parameter starts to decrease, and becomes comparable to the mass of
$\Phi$ at some point. Then $\Phi$ starts to oscillate in the potential
given as Eq.~(\ref{eq:phi_pote}). However, if the sign of $K$ is
negative, it experiences spatial instabilities, leading to the
$Q$-ball formation~\cite{Kusenko:1997si,Enqvist:1998en,Kasuya:1999wu}. 
Here we write down the equation that dictates the
configuration of the $Q$-ball and derive the solution of the
gravity-mediation type $Q$-ball.

The $Q$-ball configuration is the solution that minimizes the energy
$E$ at fixed $U(1)$ charge $Q$, where $E$ and $Q$ are defined as
\bea
E &=& \int d^3 x \left\{\sum_i \left(|\dot{\Phi}_i|^2 +|\nabla \Phi_i|^2\right) 
+V(\left\{\Phi_i\right\})\right\} ,\non\\
Q&=&-i \int d^3x \sum_i q_i \left(\Phi_i^* \dot{\Phi}_i - \dot{\Phi}_i^* \Phi_i \right).
\eea
In the single-field parametrization, they become
\bea
E&=& \int d^3 x \left\{|\dot{\Phi}|^2 +|\nabla \Phi|^2 +V(\Phi)\right\},\non\\
Q&=&-i \int d^3x~ q \left(\Phi^* \dot{\Phi} - \dot{\Phi}^* \Phi \right).
\eea
Here $q_i$ and $q$ are the $U(1)$ charges of $\Phi_i$ and $\Phi$,
respectively, and they are related to each other as
\beq
q = \frac{1}{N} \sum_i q_i.
\eeq
With the use of the method of Lagrange multipliers, the problem is reduced to
minimizing 
\bea
{\cal E}_\omega &=& \int d^3 x \left\{|\dot{\Phi}|^2 +|\nabla \Phi|^2 +V(\Phi)\right\}\non\\
&&+\omega \left[Q+i \int d^3x~ q \left(\Phi^* \dot{\Phi} - \dot{\Phi}^* \Phi \right)\right],
\eea
where $\omega$ is a Lagrange multiplier. The solution takes the
spherically symmetric form $\Phi(r,t)=
\phi(r)\, e^{i q \omega t} /\sqrt{2}$, where $\phi(r)$ satisfies the following equation,
\beq
\label{eq:qb_eq}
\frac{d^2 \phi}{dr^2}+\frac{2}{r} \frac{d \phi}{dr}-\frac{\del V_\omega(\phi)}{\del \phi}=0
\eeq
with the boundary conditions
\beq
\label{eq:bc}
\left.\frac{d \phi}{dr}\right|_{r=0}=0,~~~\phi(\infty)=0.
\eeq
Here we defined $V_\omega(\phi)=V(\phi)-q^2 \omega^2\phi^2/2$. For the
solution to exist, $V(\phi)/\phi^2$ must have a global minimum away
from the origin.

Let us consider the case that the potential is given by
Eq.~(\ref{eq:phi_pote}).  The $Q$-ball equation, Eq.~(\ref{eq:qb_eq}),
then reads
\bea
&&\frac{d^2 \phi}{dr^2}+\frac{2}{r} \frac{d \phi}{dr}+q^2\omega^2\phi \non\\
&&-m_{\Phi}^2 \phi \left(1+K+K \log\left(\frac{\phi^2}{2 N M_*^2}
\right)\right)=0.
\eea
If we adopt the Gaussian ansatz, $\phi(r)=\phi(0) e^{-r^2/R_Q^2}$, 
the radius and angular velocity of the gravity-mediation type $Q$-ball
are determined as~\cite{Enqvist:1998en}
\beq
\label{eq:3drad}
R_Q = \frac{\sqrt{2}}{\sqrt{|K|} m_\Phi}, ~~~q\,\omega = m_\Phi \sqrt{1+2|K|},
\eeq
where we set $\sqrt{N} M_*=\phi(0)/\sqrt{2}$.  The energy per unit
charge, $E/Q$, can be calculated with use of this solution:
\bea
\frac{E}{Q} &=& \frac{m_\Phi}{q} \left(1+\frac{5}{2}|K|\right)(1+2|K|)^{-1/2},\non\\
&\simeq&\frac{m_\Phi}{q} \left(1+\frac{3}{2}|K|\right),
\eea
where we assumed $|K| \ll 1$.  Note that these results depend on the
dimension of space, $D=3$.  For later use, we write down the result in
the case of $D=1$:
\bea
\label{eq:1drad}
&&R_Q = \frac{\sqrt{2}}{\sqrt{|K|} m_\Phi}, ~~~q\,\omega = m_\Phi,\non\\
&&\frac{E}{Q}= \frac{m_\Phi}{q} \left(1+\frac{|K|}{2}\right).
\eea

\section{model} 
\label{sec:model}

Now let us consider the evolution of $Q$-balls, assuming that some of
the constituent fields decay into something else. For a concrete
discussion, we take up the simplest possible flat direction composed
of two scalar fields with $U(1)$ gauge symmetry,
$X=\Phi_1(+1)\,\Phi_2(-1)$, but it is trivial to extend our results to
more generic case.  We take the global $U(1)$ charges of $\Phi_1$
and $\Phi_2$ are $q_1=q_2=q=1$~\footnote{
Even if $\Phi_1$ and $\Phi_2$ are
charged under different global $U(1)$ symmetries (e.g., the baryon and
lepton symmetries), the system is reduced to that with single global 
$U(1)$ symmetry (e.g., $U(1)_{B+L}$), as long as the A-term is given by some powers of $X$. }.
Also we assume that only
$\Phi_2$ decay into light species.  After $X$ starts to oscillate, it
feels spatial instabilities and deforms into $Q$-balls. Once $Q$-balls
are formed, the evolution of the scalar fields inside them decouple
from the cosmic expansion, so we neglect the effect of the expanding
universe in what follows.

The constituent fields, $\Phi_{1,2}$, obey the equations of
motion~\footnote{ The decay of the $Q$-ball occurs only around its
surface due to the Pauli blocking~\cite{Cohen:1986ct}, if the decay
particles are fermions. Therefore the empirical treatment of inserting
$\Gamma \dot{\Phi}_2$ might not be valid in the light of microphysical
processes. However, what we are concerned here is not the decay
process but the final state after the decay process
completed. Therefore we believe such a simplification does not spoil
our discussions.  },
\bea
\label{eq:EOM}
\ddot{\Phi}_1+\frac{\del V(\Phi_1,\Phi_2)}{\del \Phi_1^*} &=&0,\non\\
\ddot{\Phi}_2+\Gamma \dot{\Phi}_2+\frac{\del V(\Phi_1,\Phi_2)}{\del \Phi_2^*} &=&0
\eea
with
\bea
\label{eq:potential12}
V(\Phi_1,\Phi_2) &=& \sum_{i=1,2} m_{i}^2 |\Phi_i|^2 \left(
1+K_i \log\left(\frac{|\Phi_i|^2}{M_*^2}\right)\right)\non\\
&&+
\frac{g^2}{2} \left(
|\Phi_1|^2- |\Phi_2|^2\right)^2,
\eea
where $\Gamma$ is the decay
 rate of $\Phi_2$, $g$ the $U(1)$ gauge coupling
constant. Before the decay becomes effective, the $Q$-ball solution takes the
form
\beq
\label{eq:initial}
\Phi_1^{(i)} = \Phi_2^{(i)} = \frac{\Phi}{\sqrt{2}} = \frac{\phi(0)}{2}  e^{i \omega t-r^2/R_Q^2}
\eeq
with $R_Q$ and $\omega$ given in Eq.~(\ref{eq:3drad}) (or
(\ref{eq:1drad})), and $\phi(0)=2M_*$.  If we observe the $D$-term
interaction between $\Phi_1$ and $\Phi_2$, the final state can be
inferred by assuming that the $Q$-ball remains to be stable during the
decay, which will turn out to be valid later. First of all, since only
the absolute values of $\Phi_{1,2}$ participate in the $D$-term
interaction, they can exchange the energy but not the global charge.
Second, the energy of $\Phi_1$ cannot be extracted without extracting
its charge, because the motion of $\Phi_1$ is a circular orbit around
the origin.  Therefore, $\Phi_1$ will continue moving in a circular
orbit, while the angular velocity of $\Phi_2$ decreases to zero with
the $D$-flat condition, $|\Phi_2| = |\Phi_1|$, satisfied.  The final
state can be thus expressed as
\bea
\label{eq:final}
\Phi_1^{(f)} &=& \frac{\tilde{\phi}(0)}{2}  e^{i \tilde{\omega} t-r^2/\tilde{R}_Q^2},\non\\
\Phi_2^{(f)} &=& |\Phi_1^{(f)}| =  \frac{\tilde{\phi}(0)}{2}  e^{-r^2/\tilde{R}_Q^2}.
\eea
Note that the amplitude, radius, and angular velocity,
$\tilde{\phi(0)}$, $\tilde{R}_Q$, and $\tilde{\omega}$, do not
necessarily coincide with $\phi(0)$, $R_Q$, and $\omega$,
respectively.

Let us derive the $Q$-ball solution in the final state.
From the above discussion, we would like to find the solution of the
following form,
\beq
\Phi_1 = \frac{\Phi}{\sqrt{2}},~~~\Phi_2 = |\Phi_1|
\eeq
with $\dot{\Phi}_2=0$.  It should be noted that $\Phi_2$ still has the
gradient and potential energies in order to satisfy the $D$-flat
condition. The energy and charge of the system are given as
\bea
E&=& \int d^3 x \left\{\frac{1}{2}|\dot{\Phi}|^2 +|\nabla \Phi|^2\right. \non\\
&&\left.+ m_{\Phi}^2 |\Phi|^2 \left(
1+K \log\left(\frac{|\Phi|^2}{2 M_*^2}\right)\right)\right\},\non\\
Q&=&\frac{-i}{2} \int d^3x~ \left(\Phi^* \dot{\Phi} - \dot{\Phi}^* \Phi \right),
\eea
where we defined
\beq
m_\Phi^2 \equiv \frac{m_1^2+m_2^2}{2},~~~K\equiv\frac{m_1^2K_1
+m_2^2 K_2}{m_1^2+m_2^2}.
\eeq
As usual, we find the solution that minimizes 
\bea
{\cal E}_\omega &=& E+\tilde{\omega}
 \left[Q+\frac{i}{2} \int d^3x~ \left(\Phi^* \dot{\Phi} - \dot{\Phi}^* \Phi \right)\right],
\eea
where $\tilde{\omega}$ is the Lagrange multiplier.  The solution is
given as $\Phi(r,t)=\phi(r) e^{i \tilde{\omega}t}/\sqrt{2}$, where
$\phi(r)$ satisfies
\bea
\label{eq:qb_eq2}
&&\frac{d^2 \phi}{dr^2}+\frac{2}{r} \frac{d \phi}{dr}+\frac{\tilde{\omega}^2}{2}
\phi \non\\
&&-m_{\Phi}^2 \phi \left(1+K+K \log\left(\frac{\phi^2}{4 M_*^2}
\right)\right)=0
\eea
with the boundary condition given as Eq.~(\ref{eq:bc}).  Assuming the
Gaussian ansatz, $\phi(r)=\tilde{\phi}(0) e^{-r^2/\tilde{R}_Q^2}$, we
obtain
\bea
\label{eq:3drad2}
\tilde{R}_Q &=& \frac{\sqrt{2}}{\sqrt{|K|} m_\Phi},\non\\
\tilde{\omega}^2 &=& 2m_\Phi^2 \left\{1+2|K|\left(1- \log\left[\frac{\tilde{\phi}(0)}{\phi(0)}\right]\right)\right\},
\eea
Meanwhile, the charge conservation of $\Phi_1$ leads to
\beq
\omega\, \phi(0)^2 = \tilde{\omega}\, \tilde{\phi}(0)^2.
\eeq
Assuming $|K|\ll 1$, $\tilde{\omega}$ and $\tilde{\phi}(0)$ 
can be iteratively determined as
\bea
\tilde{\omega} &\simeq&(2+4|K|+|K| \log2)^{1/2} m_\Phi,\non\\
\tilde{\phi}(0) &\simeq&(2+4|K|+|K| \log2)^{-1/4} \phi(0),
\eea
The energy per unit charge, $E/Q$, can be calculated with use of this
solution, and given by
\beq
\frac{E}{Q} \simeq \sqrt{2}m_\Phi \left(1+\frac{3}{2} |K|+\frac{|K|}{4} \log{2}\right).
\eeq
Thus the $Q$-ball configuration after the decay of $\Phi_2$
has the same shape in the real space, but the scalar field inside the
$Q$-ball moves in a different orbit.
We write down the similar results in the case of $D=1$.
\bea
\label{eq:1drad2}
\tilde{R}_Q &=& \frac{\sqrt{2}}{\sqrt{|K|} m_\Phi}, \non\\
\tilde{\omega} &\simeq& \left(2+|K| \log2\right)^{1/2} m_\Phi,\non\\
\tilde{\phi}(0) &\simeq&\left(2+|K| \log2\right)^{-1/4} \phi(0), \non\\
\frac{E}{Q} &\simeq& \sqrt{2}\left(1+\frac{|K|}{2} +\frac{|K|}{4}\log{2}\right)m_\Phi.
\eea
\begin{figure}[t]
\includegraphics[width=6cm]{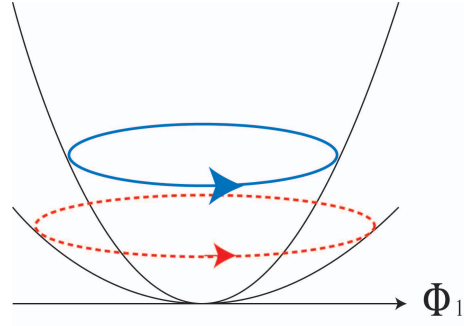}
\caption{
The schema of the effective potential of $\Phi_1$. 
Before $\Phi_2$ decays,
$\Phi_1$ inside the $Q$-ball moves in a circular orbit 
shown as the dotted line. After the decay of $\Phi_2$, 
the amplitude of $\Phi_1$ decreases 
but its angular velocity increases so that the charge 
of $\Phi_1$ remains constant. The modified orbit is 
represented as the solid circle.
}
\label{fig:insideQ}
\end{figure}

We would like to make several remarks. First, the $Q$-ball solution
such as Eq.~(\ref{eq:final}) exists if and only if the $Q$-ball
solution is allowed before the decay of $\Phi_2$, that is to say,
$K<0$.  In other words, the partial decay of the constituent fields
does not split the $Q$-ball into fragments. This is rather surprising
result. For instance, even if $K_1$ is positive, the $Q$-ball
configuration remains stable as long as $K$ is negative. It is the
conservation of the charge of $\Phi_1$ that keeps the $Q$-ball stable 
during the course of the decay. After the decay of  $\Phi_2$ completed,
$\Phi_1$ still feels the potential of $\Phi_2$ through the $D$-term
potential. Therefore $\Phi_1$ moves in an effective potential given as
the sum of the potentials of $\Phi_1$ and $\Phi_2$.  See Fig.~\ref{fig:insideQ}.
Second, the energy per unit charge becomes about $\sqrt{2}$ times larger, which
means the $Q$-ball configuration becomes slightly unstable. The
resultant $Q$-ball might be able to decay into light particles by
using this increase. Such increment comes from the fact that the interplay of
the charge conservation of $\Phi_1$ and the
$D$-flat condition do not allow $\Phi_2$ to decay with the potential
energy. Third, the reason why the energy of $\Phi_1$ cannot dissipate
through the decay channel of $\Phi_2$ is that the $Q$-ball is  a
state of the lowest energy.  It is helpful to consider how the
$I$-ball~\cite{Kasuya:2002zs}, which is considered to be in an excited
state of the $Q$-ball, change its profile if the partial decay
occurs. The $I$-ball solution exists if the scalar potential is
dominated by the quadratic term, which guarantees the approximate
invariance of the adiabatic charge, $I$. The $I$-ball configuration is
obtained as a state with the lowest energy at fixed $I$. Its profile
is very similar to that of the $Q$-ball, except that the scalar field
moves in an orbit with large ellipticity.  If the kick of A-term is
not strong enough, $I$-balls are generally formed instead of
$Q$-balls.  Let us consider the $I$-ball composed of $X=\Phi_1\Phi_2$,
and suppose that $\Phi_2$ decays into something else.  In this case,
the energy of $\Phi_1$, in addition to $\Phi_2$, can dissipate into light
particles through the
$D$-term interaction. This process lasts until the ellipticity of the
orbit becomes $\sim1$, {\it i.e.}, until the $I$-ball becomes a
$Q$-ball, from which the energy cannot be extracted anymore without
violating the charge conservation of $\Phi_1$.

\section{Numerical Calculation} 
\label{sec:nume}
We have performed the numerical calculation to validate the analysis
presented in the previous section. In particular, it is important to
follow the evolution and show that the $Q$-ball does not break up
during the decay of $\Phi_2$.

We have solved the equations of motion, Eq.~(\ref{eq:EOM}), on the one
dimensional lattices with the initial condition given as
Eq.~(\ref{eq:initial}). The space $x$ and time $t$ are normalized by
a mass scale $m$, while the field value is normalized by $M_*$. We
take the following values for the model parameters: $m_1=m_2=m$,
$\phi(0)/M_*=2$, $g^2 M_*^2 = 200 m^2$, $K_1=0.02$, $K_2=-0.1$,
and $\Gamma = 0.1 m$. 

\begin{figure}
\includegraphics[width=6.5cm]{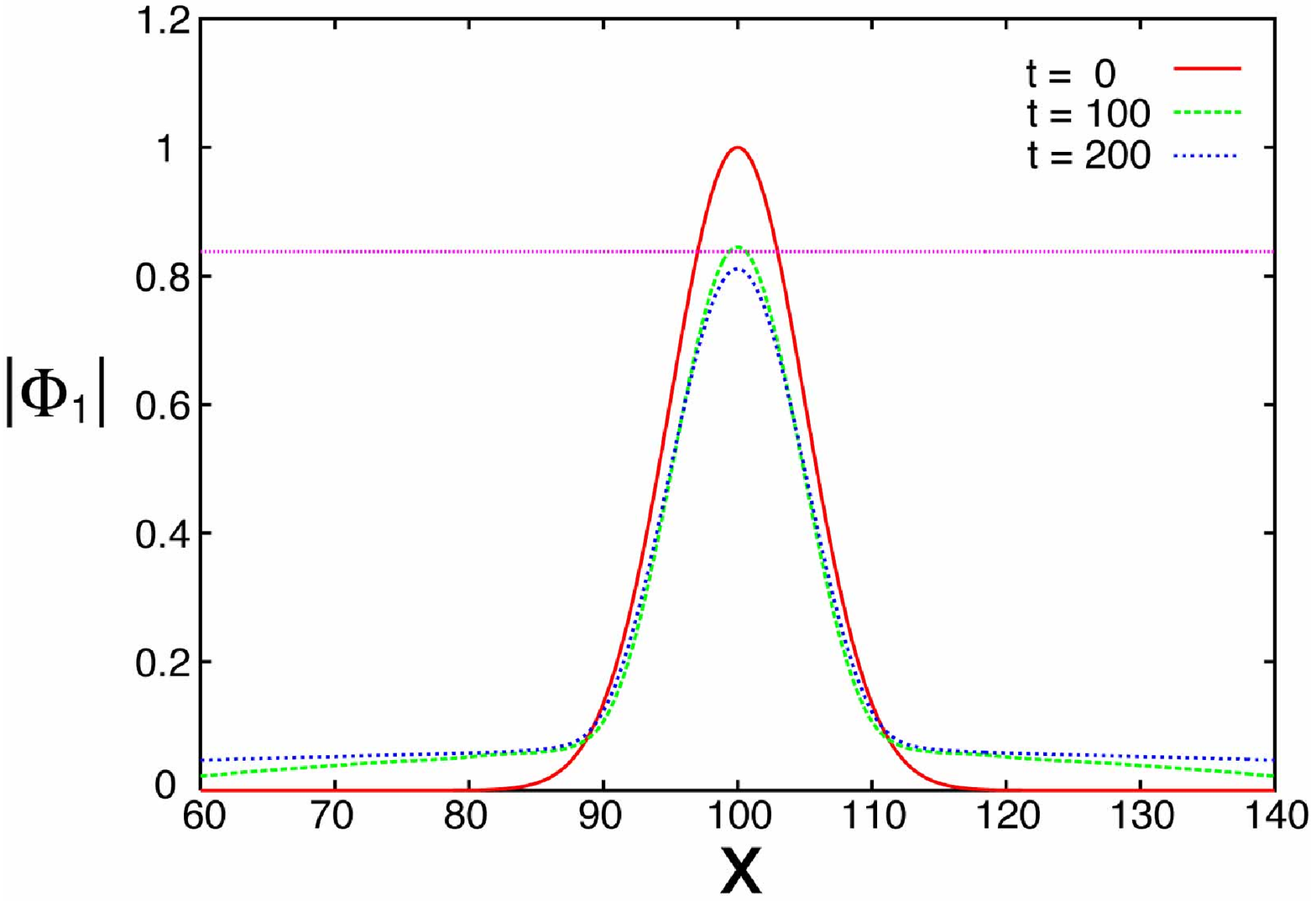}
\includegraphics[width=6.5cm]{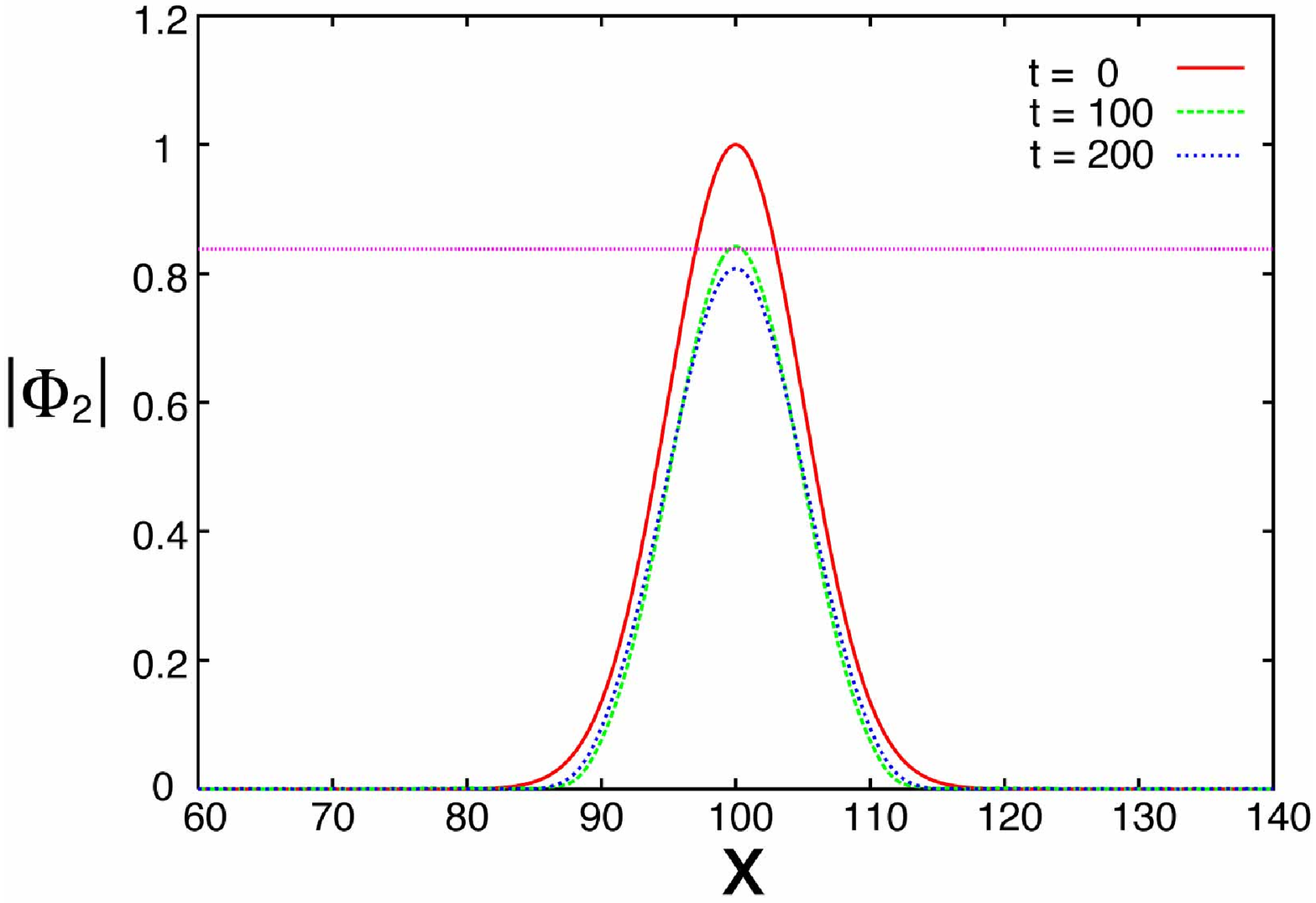}
\caption{The evolution of the distribution of $|\Phi_1|$ 
and $|\Phi_2|$. The horizontal straight line represents 
the analytical estimate of $\tilde{\phi}$ given in
Eq.~(\ref{eq:1drad2}). The $D$-flat condition is satisfied 
inside the $Q$-ball centered at $x=100$.
The span of the lattices is actually $x=0\sim200$.
}
\label{fig:amp}
\end{figure}

\begin{figure}
\includegraphics[width=6.5cm]{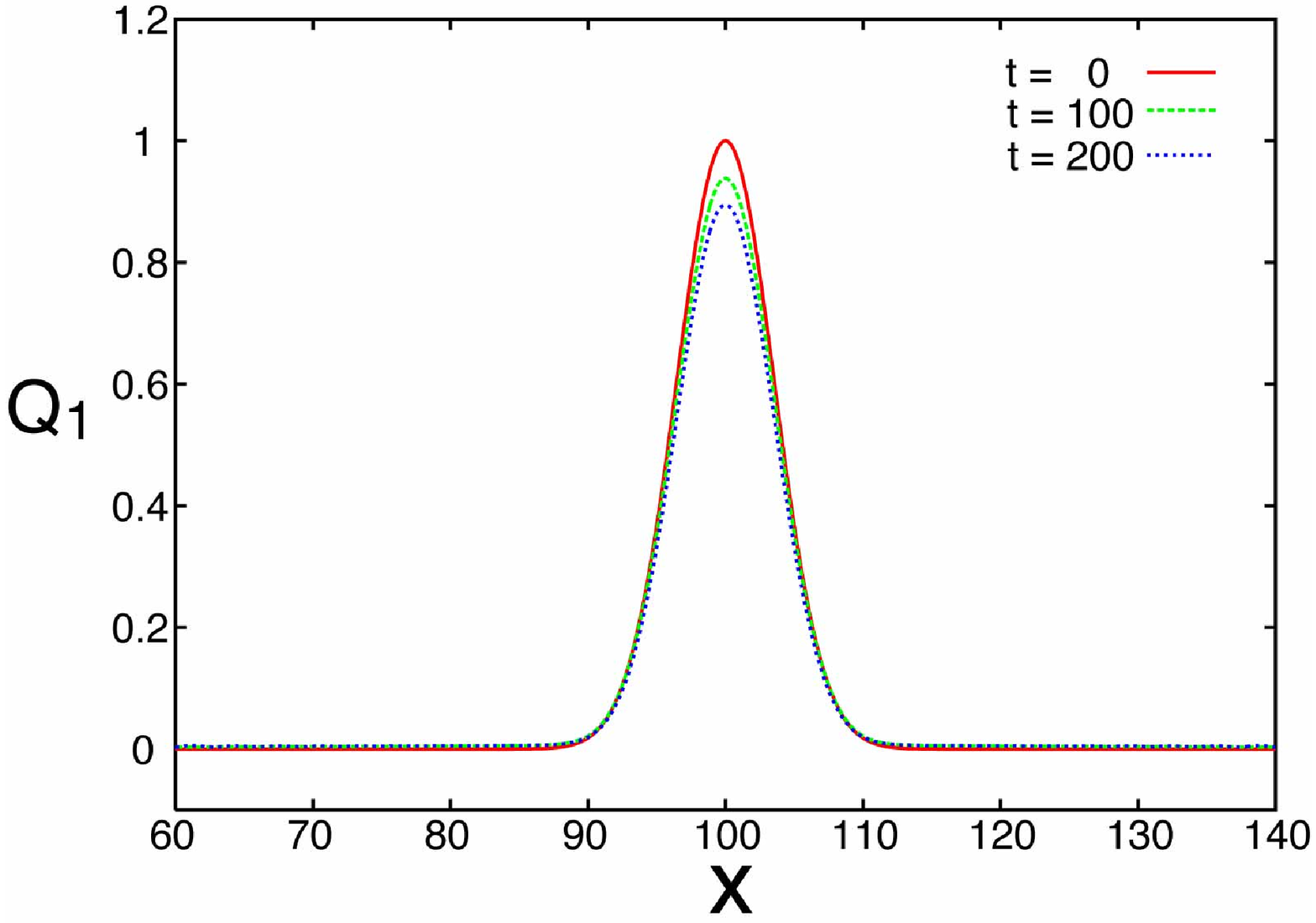}
\includegraphics[width=6.5cm]{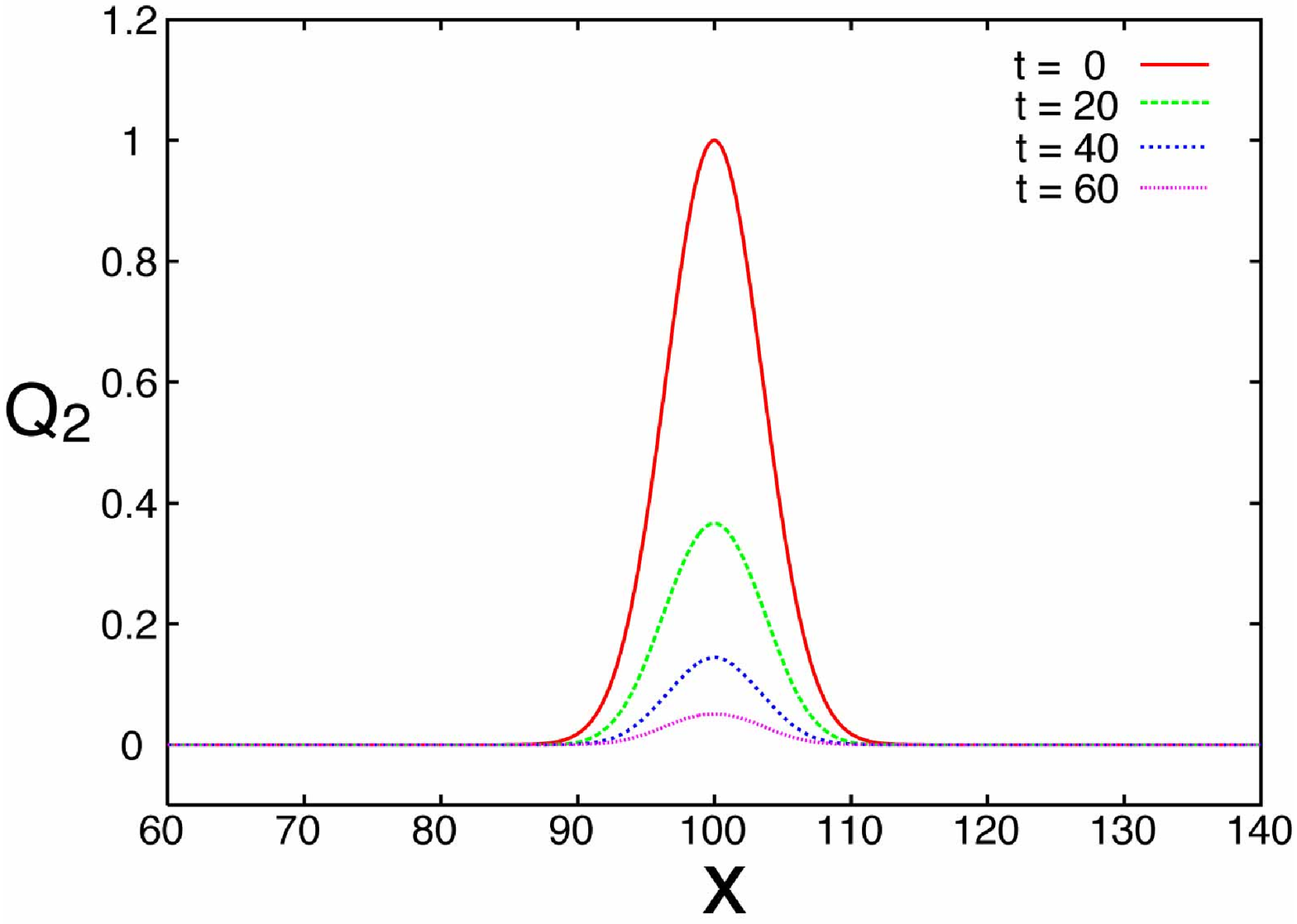}
\caption{The evolution of the charge distribution of 
$\Phi_1$ and $\Phi_2$.
While $Q_1$ is conserved, $Q_2$ rapidly decreases  to zero 
due to the decay of $\Phi_2$. The charge outside the $Q$-ball 
is negligibly small.
}
\label{fig:charge}
\end{figure}

\begin{figure}
\includegraphics[width=6.5cm]{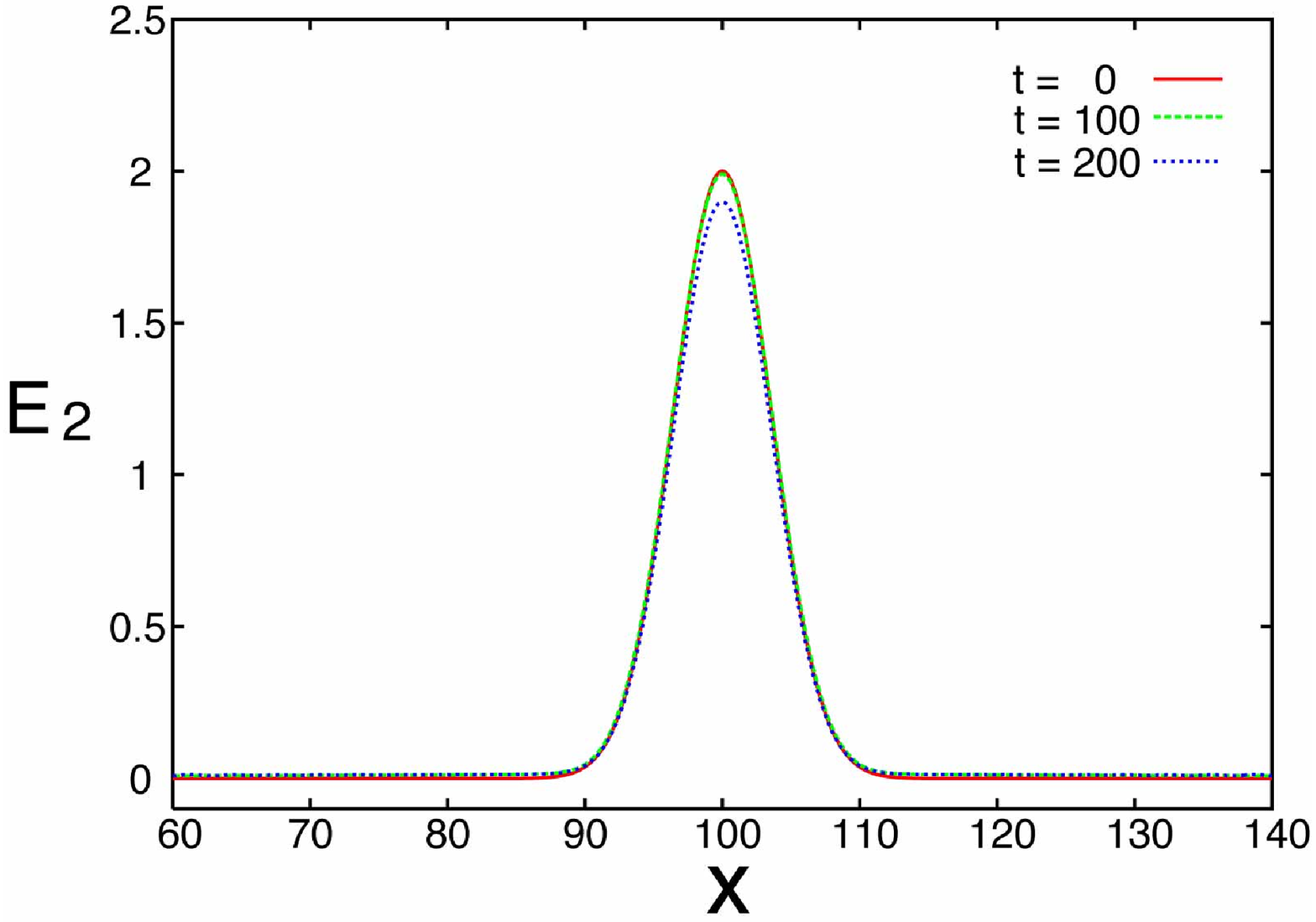}
\includegraphics[width=6.5cm]{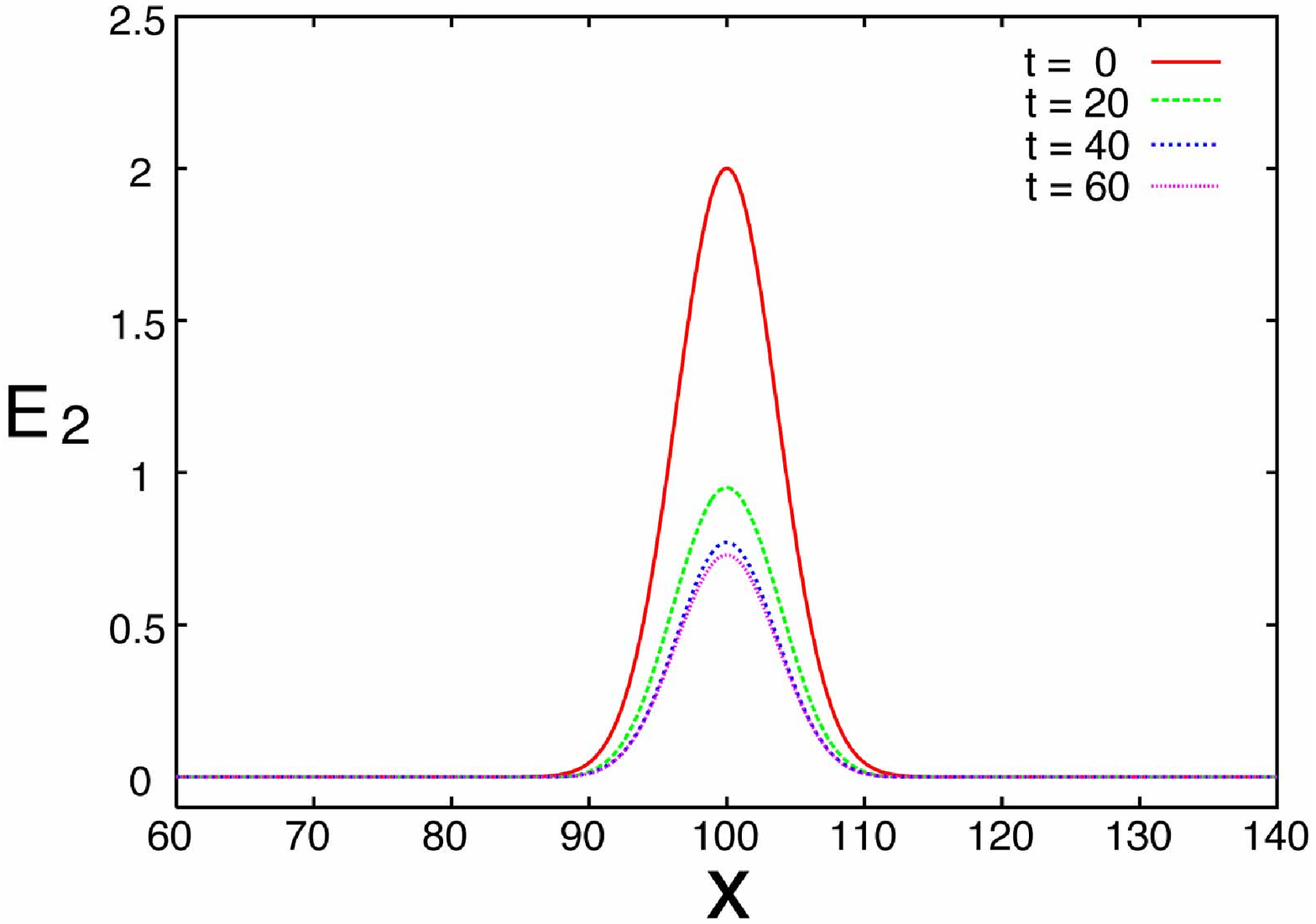}
\caption{The evolution of the energy distribution.
While $E_1$ stays unchanged, $E_2$ decreases to some finite
value, which corresponds to the potential energy of $\Phi_2$.
We do not have included the $D$-term potential 
in the definition of $E_{1,2}$.
}
\label{fig:energy}
\end{figure}

\begin{figure}[htbp]
\includegraphics[width=7cm]{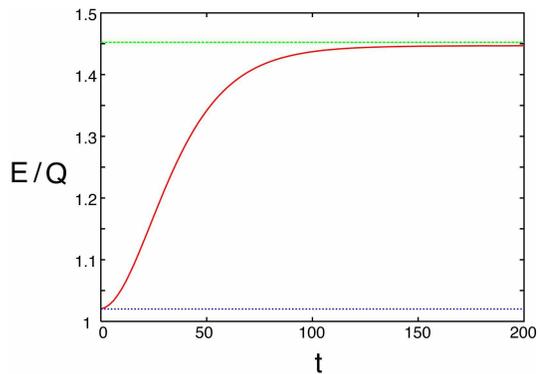}
\caption{The time evolution of the energy per unit charge, $E/Q$.
The analytical estimates for the initial and final states 
are shown as the horizontal straight lines.
}
\label{fig:EoverQ}
\end{figure}

The evolutions of $|\Phi_1|$ and $|\Phi_2|$ are
illustrated in Fig.~\ref{fig:amp}. 
As seen in the figure, the $D$-flat condition
is satisfied inside the $Q$-ball but not outside. However,  as shown
in Figs.~\ref{fig:charge} and \ref{fig:energy}, 
the leakage of the energy or charge of $\Phi_1$ from the $Q$-ball 
is negligible~\footnote{
For a long-time simulation, the leakage of the charge and energy from the
$Q$-ball cannot be negligible, as suggested by the slight decrease of
$Q_1$ and $E_1$ in Figs.~\ref{fig:charge} and  \ref{fig:energy}, respectively. 
In the realistic situation, the $D$-flat condition is much severer than in the
numerical calculation, which would highly
suppress such leakage.
}.
The amplitude decreases as
expected by the analytical estimate (see Eq.~(\ref{eq:1drad2})).
The charge and energy of $\Phi_1$, $Q_1$ and $E_1$,
do not change before and after
the decay.  See Figs.~\ref{fig:charge} and \ref{fig:energy}. On the other hand,
the charge of $\Phi_2$ quickly decreases to zero, and its energy, $E_2$,
is also brought down to the potential energy of $\Phi_2$, which is
necessary to satisfy the $D$-flat condition. 
 It should be noted that 
the shape of the $Q$-ball remains unchanged throughout the decay as expected.
The behavior of the energy per unit charge, $E/Q$, is shown in 
Fig.~\ref{fig:EoverQ}, and it agrees well with the predicted values given in
Eqs.~(\ref{eq:1drad}) and (\ref{eq:1drad2}).


\section{discussions} 
\label{sec:discuss}
So far we have presented the analytical and numerical 
study of the partially decaying $Q$-balls with use of a toy
model of flat directions. Let us consider now the implication 
of the results presented above. We have assumed that only 
a part of constituent fields decay, but our arguments apply
to more generic case where there is a hierarchy between the 
decay rates, say, $\Gamma_1 \gg \Gamma_2$. 
The asymmetry of the 
$Q$-ball continuously changes during the course of the decay.
Most of the flat directions in
the MSSM are composed of both squarks and sleptons,
therefore they have both baryon and lepton number. 
If such a flat direction deforms into $Q$-balls and they are
stable, our argument can be applied 
since the slepton condensate can always decay into a pair 
of neutrinos through the exchange of the gauginos. 
All through the decay processes, the $Q$-balls sequentially  
transform themselves toward pure $B$-balls with no lepton number. 
In the usual scenario considered
so far, it is a flat direction composed only of squarks that forms
pure $B$-balls. Therefore our analysis indicates that the pure $B$-balls
are generated for most of flat directions, as a result of the
decay into neutrinos. Specifically, all the $Q$-ball dark matter
must be pure $B$-balls.
Interestingly enough, the resultant $B$-ball might become
unstable due to the increase of its energy per unit charge,
leading to a complete decay of the $Q$-ball.
In connection with the transformation of $Q$-balls into
$B$-balls, the  leptogenesis  might become possible
even for a flat direction with $B-L =0$. 
Another interesting possibility is that the
$I$-ball  can be naturally altered to Q-ball, if one of the constituent fields
decay into something else. In particular, the $I$-ball composed
of  a flat direction that includes a leptonic field  is generally transformed
into a $Q$-ball. In fact, since the $I$-ball is a quasi-stable
object, its conversion  to  a $Q$-ball was not  known.
The results of this study affords some new perspectives on the
decay or stabilization processes of non-topological solitons
made of flat directions.

In summary, we have investigated the evolution of  $Q$-balls,
assuming that some of the constituent scalar fields decay into
light particles. With a toy model of the flat direction,
we have obtained the analytical solution of the
$Q$-ball in the final state, and found that the spatial shape
of the $Q$-ball does not change, but the orbit of the scalar field
inside the $Q$-ball does change in order to compensate for the 
decay. In particular, the energy per unit charge generally increases
due to the partial decay, which might induce further decay of
the remnant scalar fields. Also we have performed numerical calculations
to confirm these results. Again we stress that $Q$-balls remain $Q$-balls
throughout the partial decay,  due to both the charge conservation of 
the remnant scalar fields and the $D$-flat condition.

\subsection*{ACKNOWLEDGMENTS}
This work was partially supported by the JSPS Grant-in-Aid for Scientific 
Research No.\ 10975 (F.T.)

 \end{document}